\documentclass[journal]{IEEEtran}
\usepackage{booktabs}      
\usepackage{amsmath}       
\usepackage{graphicx}      
\usepackage{float}
\usepackage{orcidlink}

\hypersetup{
    hidelinks
}

\ifCLASSINFOpdf
\else
\fi
\begin{document}
%
\title{Photonic Memory as a Dynamical Phase for Neuromorphic Computing: A Unified Framework with Experimental Realization}
%
%
%

\author{Isaac Yorke, \orcidlink{0000-0002-1645-6470}\\
 Department of Engineering and Architecture, \\University of Parma, Parco Area delle Scienze 181/A I-43124 Parma, Italy\\
e-mail: isaac.yorke@unipr.it  or  yorkeisaac034@gmail.com}
%
%

\markboth{This manuscript is a preprint and has not been peer reviewed}
{Shell \MakeLowercase{\textit{et al.}}: Bare Demo of IEEEtran.cls for IEEE Journals}
%



\maketitle

\begin{abstract}
Photonic memory underpins optical information processing, neuromorphic photonics, and photonic computing. Existing studies typically treat dispersive, nonlinear, and driven–dissipative memory as distinct physical phenomena, despite all being governed by the evolution of the optical field. This work proposes a unified phase-based framework in which photonic memory is interpreted as a dynamical phase, with transitions between memory regimes governed by optical phase evolution, Kerr nonlinearity, and the balance between delayed feedback and dissipation. Within this framework, dispersive memory arises from frequency-dependent phase accumulation, nonlinear memory emerges through intensity-dependent phase evolution leading to bistability and hysteresis, and driven–dissipative memory is established through attractor convergence and memory stabilization. The proposed framework is validated through theoretical analysis and numerical simulations, demonstrating a continuous progression from linear dispersive memory to nonlinear and ultimately driven–dissipative memory. Experimental validation is performed using a silicon photonic waveguide incorporating chirped Bragg gratings. Group delay measurements reveal distinct linear and nonlinear memory responses, while the reconstructed memory distribution demonstrates the coexistence of dispersive, nonlinear, and driven–dissipative memory within a single integrated photonic device. To the best of the author's knowledge, this constitutes the first experimental demonstration supporting the coexistence of all three photonic memory regimes in a single photonic platform. These results establish optical phase as the unifying physical quantity underlying photonic memory and provide a common framework for designing future neuromorphic photonic systems, reservoir computers, and integrated photonic processors.
\end{abstract}


%
\IEEEpeerreviewmaketitle

\section{Introduction}
%
%
%
%
\IEEEPARstart{M}{emory} is one of the fundamental requirements for neuromorphic computation \cite{goi2020perspective}, and different photonic memory mechanisms have traditionally been treated separately \cite{farmakidis2024integrated}. A unified framework provides a common language for understanding and designing photonic neuromorphic systems. Photonic memory has emerged as a key concept in optical communications, neuromorphic photonics, reservoir computing, and photonic information processing \cite{goi2020perspective, van2017advances, yorke2026reconfigurable}. Unlike conventional electronic memory, which is commonly implemented using static storage elements that retain information through stable physical states, photonic memory arises from dynamical processes that enable optical systems to retain information about previous states \cite{patterson2022computer, prucnal2017neuromorphic}. Such memory effects have been observed across a wide range of photonic platforms, including dispersive waveguides supporting fading-memory dynamics \cite{van2017advances}, nonlinear optical resonators exhibiting bistability and hysteresis \cite{gibbs2012optical}, and driven–dissipative photonic systems characterized by attractor-based state retention \cite{lukovsevivcius2009reservoir}. Despite their apparent differences, these mechanisms share a common feature: the retention of information through the evolution of the optical field.
Existing studies generally investigate dispersive, nonlinear, and driven–dissipative memory mechanisms within separate theoretical and experimental frameworks, emphasizing their specific physical origins rather than a common dynamical principle \cite{paparelle2026experimental, castro2024memory, labay2023quantum}. In dispersive systems, memory emerges from the wavelength-dependent accumulation of phase, which produces pulse broadening and temporal fading. In nonlinear systems, memory arises from intensity-dependent phase modulation and nonlinear state transitions such as optical bistability and hysteresis. In driven–dissipative systems, memory is associated with the convergence of system trajectories toward stable attractors maintained by the balance between external driving and dissipation. While these mechanisms are often investigated independently, a unified physical interpretation remains lacking.
This work proposes that photonic memory can be viewed as a dynamical phase whose manifestation depends on the dominant physical process governing the evolution of the optical field. The central quantity underlying this framework is the optical phase, represented by the complex optical field: \begin{equation}
E(z,t)=A(z,t)\exp\!\left(i\phi(z,t)\right),
\end{equation}
where $A(z,t)$ is the field amplitude and $\Phi(z,t)$ is the optical phase. The phase evolution of the optical field provides a natural dynamical variable through which memory formation can be understood across a wide range of photonic and neuromorphic systems \cite{bea1991fundamentals}. This work argues that dispersive, nonlinear, and driven–dissipative memory regimes can all be understood as distinct manifestations of phase evolution. Transitions between these regimes are governed by the interplay between phase accumulation, nonlinear phase modulation, and the balance between driving and dissipation.
To establish this framework, first, the dispersive memory regime is  investigated through simulations of pulse broadening and temporal fading. Then the nonlinear memory regime is examined using optical bistability and hysteresis, followed by the driven–dissipative memory regime through attractor convergence dynamics. Together, these examples demonstrate that memory behavior can emerge from fundamentally different photonic processes while remaining connected through the common language of phase evolution.
Finally, the proposed framework is experimentally realized using a silicon photonic waveguide incorporating chirped Bragg gratings. By analyzing the measured group delay response, it is shown that signatures associated with dispersive, nonlinear, and driven–dissipative memory behaviors can coexist within a single photonic device. The experimental results suggest that memory should not be regarded as a unique physical mechanism, but rather as a dynamical phase whose observable characteristics depend on the dominant operating regime of the photonic system. The novelty of this work does not lie in the individual physical mechanisms of dispersive, nonlinear, or driven–dissipative photonic memory, each of which has been studied extensively in the literature. Rather, the principal contribution is the development of a unified dynamical framework in which these apparently distinct memory mechanisms are interpreted as different manifestations of optical phase evolution. To the best of the author's knowledge, this work further provides the first experimental evidence consistent with this unified interpretation within a single silicon photonic device.

\section{Unified Phase-Based Framework for Photonic Memory }
This study has hinted that dispersion, nonlinearity, and driven–dissipative systems are all fundamentally governed by $\Phi(z,t)$, the evolution of optical phase. This provides the basis for a unified phase-based framework in which photonic memory is viewed as an emergent property of phase dynamics. Under this perspective, different memory modalities correspond to distinct manifestations of a common underlying phase evolution process. Starting from the dispersive wave equation,  $\frac{\partial A}{\partial z} = -\frac{i\beta_2}{2}\frac{\partial^2 A}{\partial t^2}$, where $A(z,t)$ is the optical field envelope, $z$ is the propagation distance, $t$ is the time and $\beta_2$ is the group-velocity dispersion parameter (GVD), dispersive memory can be interpreted as a consequence of phase evolution. This equation represents the dispersion-only form of the linear Schrödinger equation \cite{agrawal2012nonlinear} and shows that the optical phase evolves according to the spectral content of the field. As a result, information from previous temporal states is retained through dispersion-induced phase correlations, giving rise to memory. The physical interpretation of this equation is that dispersion induces frequency-dependent phase evolution, resulting in temporal broadening of the optical pulse. As a consequence, information from earlier temporal states remains embedded within the evolving field and influences subsequent states through temporal correlations. Memory therefore emerges through the persistence of phase information over time. By introducing the Kerr nonlinear term, $i\gamma |A|^2 A$, into the linear Schrödinger equation, the nonlinear Schrödinger equation (NLSE) is obtained as 
\begin{equation}
\frac{\partial A}{\partial z} = -\frac{i\beta_2}{2}\frac{\partial^2 A}{\partial t^2}+i\gamma |A|^2 A
\label{eq:NLSE}
\end{equation}
where $\gamma$ is the nonlinear Kerr coefficient \cite{agrawal2012nonlinear, ablowitz1981solitons}. The nonlinear phase accumulation per unit length is given by $\Phi_{NL} = \gamma |A|^2$. The physical interpretation is that the nonlinear Kerr effect induces intensity-dependent phase evolution. Consequently, high-intensity regions accumulate more phase, while low-intensity regions accumulate less phase. This leads to spectral broadening, self-phase modulation, nonlinear feedback, hysteresis, and bistability \cite{agrawal2012nonlinear, saleh2019fundamentals}. The nonlinear Schrödinger equation describes spatial evolution along the propagation coordinate $z$, whereas driven–dissipative systems are naturally formulated in temporal evolution $t$. Despite this difference, both can be written in the unified form of a first-order evolution equation for a complex field envelope, thanks to the Lugiato–Lefever equation \cite{chembo2013spatiotemporal, coen2012modeling, herr2016dissipative}. The unified photonic memory dynamics can therefore be expressed within a driven–dissipative Kerr cavity framework described by the Lugiato–Lefever equation:
\begin{equation}
\frac{\partial A(t,\tau)}{\partial t} = [-\alpha - i\delta] A + i\gamma |A|^2 A + i\frac{\beta_2}{2} \frac{\partial^2 A}{\partial \tau^2} + S
\label{eq:LLE}
\end{equation}
where $A(t,\tau)$ is the slowly varying field envelope, $t$ denotes the slow evolution time, and $\tau$ is the fast time in the moving frame. The term $i\frac{\beta_2}{2} \frac{\partial^2 A}{\partial \tau^2}$ captures dispersive temporal coupling, $i\gamma |A|^2 A$ represents Kerr-induced nonlinear phase evolution, $-\alpha A$ accounts for cavity losses, $-i\delta A$ describes detuning-induced phase rotation, and $S$ is the external driving field \cite{chembo2013spatiotemporal, lugiato1987spatial}. In this formulation, photonic memory emerges from the interplay between dispersion, nonlinearity, dissipation, and continuous driving within a single dynamical phase evolution framework. This work therefore proposes that $ \text{Memory} = \text{a dynamical phase of photonic systems} $,  whose manifestation depends on the dominant physical mechanism. Thus; 
\[ M = M_{disp} + M_{NL} + M_{DD} \]

\section{Dispersive Photonic Memory Regime }
\begin{figure}[!htbp]
    \centering
    \includegraphics[width=0.45\textwidth]{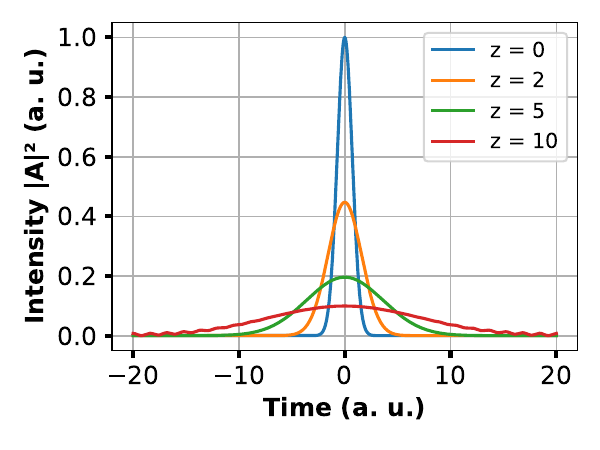}
    \caption{The temporal spreading of a pulse which creates correlations between earlier and later temporal states, confirming dispersive memory via temporal spreading.}
    \label{fig:pulse spreading}
\end{figure}

In this section, the goal is to numerically demonstrate that 
\begin{equation}
\frac{\partial A}{\partial z}=-i\frac{\beta_2}{2} \frac{\partial^2 A}{\partial t^2}
\label{eq:linear_dipersion}
\end{equation}
produces temporal spreading (thus; pulse broadening with propagation) and fading temporal correlations (thus; the memory of the initial pulse gradually becomes distributed and weakened). The Fourier transform of equation \ref{eq:linear_dipersion} with respect to $t$ can be written as: $\mathcal{F}\left\{ \frac{\partial^2 A}{\partial t^2} \right\}=-\omega^2 \tilde{A}$, which gives 
\begin{equation}
\frac{\partial \tilde{A}}{\partial z}=-i\frac{\beta_2}{2}(-\omega^2)\tilde{A} 
\label{eq:transform_disp_regime_eqn}
\end{equation}
Integrating equation \ref{eq:transform_disp_regime_eqn} gives:
\begin{equation}
\tilde{A}(z,\omega)=\tilde{A}(0,\omega)exp\left(i\frac{\beta_2}{2} \omega^2 z \right) 
\label{eq:solution_to_disp_eqn}
\end{equation}
\cite{agrawal2012nonlinear}. The amplitude of equation \ref{eq:solution_to_disp_eqn} is: $|\tilde{A}(z,\omega)| = |\tilde{A}(0,\omega)|$, which does not change as the signal is propagating. The phase of equation \ref{eq:solution_to_disp_eqn} is: $\Phi(z, \omega)=(\frac{\beta_2}{2} \omega^2 z)$, which varies as the signal is propagating. This demonstrate that dispersion does not create memory through amplitude storage; it creates memory through frequency-dependent phase accumulation. Thus; the phase accumulated through dispersion: $\Phi_{disp} \propto \beta_2 \omega^2 z$. In the simulation to demonstrate memory via temporal spreading, the intensity of a Gaussian pulse propagating through a dispersive medium using Fourier propagation was plotted. The initial pulse is $A(0,t)=exp\left(-\frac{t^2}{2 T_{0}^2} \right) $, where $T_{0}$ is the pulse width. $\beta_{2}=1$, $T_{0}=1$, $z=[0, 2, 5, 10]$, Number of temporal samples $(Nt)=2048$, maximum simulation time window $t_{max}=20$, time grid $(t)=(-t_{max}, t_{max}, Nt)$, and temporal sampling interval $(dt)=t[1] - t[0]$. Figure \ref{fig:pulse spreading} shows the intensity versus time plot of the Gaussian pulse. It can be observed that initial pulse is narrow and it broadens with propagation distance. The temporal spreading observed in Figure \ref{fig:pulse spreading} creates correlations between earlier and later temporal states, which constitute the dispersive memory mechanism within the proposed phase-based framework.

\begin{figure}[!htbp]
    \centering
    \includegraphics[width=0.45\textwidth]{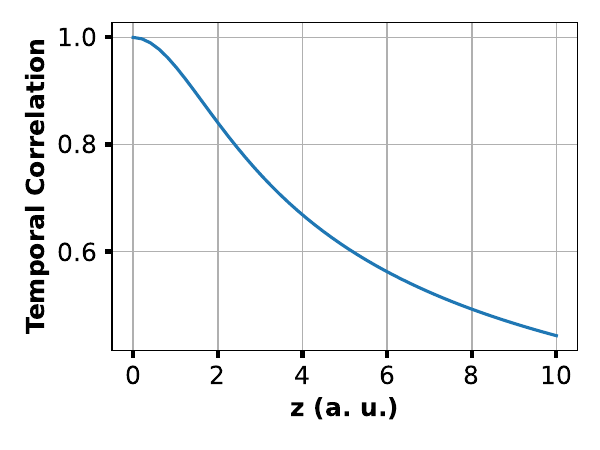}
    \caption{correlation between initial pulse and propagated pulses as a function $z$, which demonstrate memory retention loss}
    \label{fig:correlation}
\end{figure}

To demonstrate fading temporal correlation or the persistence of temporal information during propagation, the correlation $C(Z)$  between initial pulse and propagated pulses was computed using the normalized field overlap: $C(Z)=\frac{\int A^*(0,t) A(z,t) \, dt}{\sqrt{\int |A(0,t)|^2 \, dt \int |A(z,t)|^2 \, dt}}$ \cite{saleh2019fundamentals, asuero2006correlation}. Where $^*$ denotes complex conjugation. Figure \ref{fig:correlation} shows the plot of temporal correlation as a function of $z$, which reveals that correlation between initial pulse and propagated pulses starts near $1$, indicating identical fields and gradually decreases with $z$. This confirms fading temporal correlation and memory retention loss.

Figures \ref{fig:pulse spreading} and \ref{fig:correlation} demonstrate that, dispersion causes temporal information to spread, earlier states to persist, but correlations gradually fade. Thus: dispersion acts as a distributed fading-memory operator.

\section{Nonlinear Photonic Memory Regime  }
\begin{figure}[!htbp]
    \centering
    \includegraphics[width=0.45\textwidth]{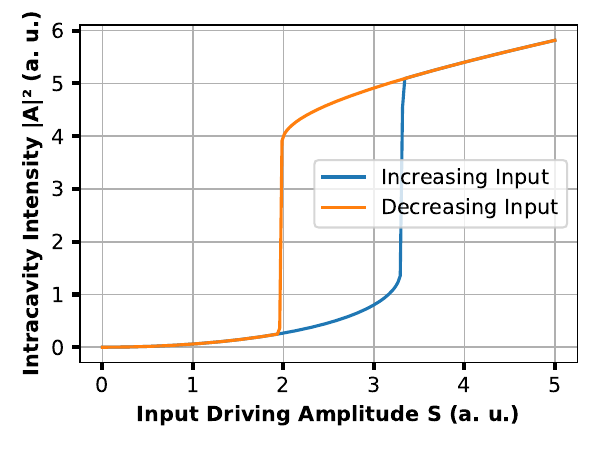}
    \caption{Optical bistability and hysteretic memory formation in the nonlinear Kerr regime. The nonlinear Kerr regime exhibits optical bistability and hysteresis arising from intensity-dependent phase evolution. The existence of distinct switching thresholds during increasing and decreasing input sweeps demonstrates history-dependent state retention and nonlinear photonic memory formation}
    \label{fig:bistability_hysteresis}
\end{figure}

Equation \ref{eq:NLSE} displays the nonlinear Schrödinger equation, where the Kerr nonlinearity factor $i\gamma|A|^2 A$ is introduced. The NLSE describes a pulse propagating through a Kerr medium, and the Kerr term produces the nonlinear phase $\Phi_{NL}=\gamma|A|z$ \cite{loures2015contribution}. This leads to self-phase modulation (SPM), spectral broadening, and self-trapping, soliton formation (when balanced by dispersion) \cite{agrawal2012nonlinear, loures2015contribution, mezache2026nonlinear}. However, the field experiences the nonlinearity only once as it propagates through the medium. There is no mechanism for repeated interaction with previous field states, so bistability and hysteresis generally do not arise in pure single-pass propagation \cite{maurya2012pattern, lidorikis1999wave}. To observe hysteresis and bistability, it is imperative to transition from traveling-wave Kerr propagation to cavity-enhanced Kerr dynamics, which is where optical bistability, hysteresis, and memory effects will emerge. Therefore, the reduced  form of equation \ref{eq:LLE} is introduced, which is referred to as; reduced driven–dissipative Kerr cavity model: 
\begin{equation}
    \frac{dA}{dt}=-(1+i\delta)A+i\gamma|A|^2 A+S
\label{eq:reduced_LLE}
\end{equation}
where the parameters have the same meaning as in equation \ref{eq:LLE}. Equation \ref{eq:reduced_LLE} tells us that the cavity introduces a feedback mechanism whereby a fraction of the optical field remains stored and recirculates after each round trip. Consequently, the intracavity field depends on its previous states, providing the physical basis for nonlinear photonic memory. This feedback mechanism is the origin of optical bistability, hysteresis and nonlinear photonic memory in Kerr cavities. $-A$ represents cavity decay (loss of stored energy) whilst $\delta$ is the detuning parameter and it measures the frequency (or wavelength) mismatch between the driving laser and the cavity resonance. In the reduced model, the term $-i\delta A$ produces a phase rotation of the intracavity field. Because the Kerr effect changes the refractive index, the Kerr phase shift depends on intensity, the resonance condition shifts dynamically \cite{priem2003optical}, and feedback creates multiple stable branches \cite{petracek2014simulation}. To demonstrate bistability and hysteresis, since in the nonequilibrium
steady-state regime, $dA/dt=0$ \cite{yorke2026reconfigurable},  the steady-state dynamics of equation \ref{eq:reduced_LLE} is solved, by slowly varying $S$ and tracking $|A|^2$. The parameters were set as: $S=[0,1,2,3,4,5]$, $\delta=4$, $\gamma=1$ and  $dt=0.01$. Figure ~\ref{fig:bistability_hysteresis} shows the plot of the intracavity intensity $|A|^2$ as a function of external driving amplitude $S$. In Figure ~\ref{fig:bistability_hysteresis}, there are two curves. For the increasing input (blue curve), as input $S$ increases, intracavity intensity increases gradually, until a critical threshold, then the system suddenly jumps to a high-intensity branch. This jump occurs because: Kerr nonlinearity shifts the resonance, feedback reinforces the field to cause nonlinear self-amplification. This is nonlinear switching. For the decreasing input (orange curve), as input starts to decrease,
the system remains on the high branch. This means the system remembers past state, even below the original switching threshold. Only at a much lower input does it collapse back. This means the current state depends on history, which is a demonstration of hysteretic memory. Between $S \approx 2$  and  $S \approx 3.3$, there are two possible stable states for the same input. Thus; low-intensity state exists and high-intensity state also exists. Which state the system occupies depends on previous evolution and this is a demonstration of optical bistability. Overall, Figure ~\ref{fig:bistability_hysteresis} demonstrates that The system retains its previous branch, it resists immediate switching, and preserves state information. Thus, memory emerges from intensity-dependent phase evolution, which enables the retention of information about previous optical states. The system therefore exhibits key signatures of nonlinear photonic memory, neuromorphic photonics, and driven nonlinear systems. The system also demonstrates self-trapping because, once the it enters the high branch, it remains localized there, despite decreasing input. This nonlinear photonic memory, which shows stable branch retention, history dependence, and nonlinear persistence confirms the definition of memory in \cite{yorke2026reconfigurable}, which is; memory forms when weights remain in stable hysteretic branches despite fluctuations.

\section{Driven-Dissipative Photonic Memory Regime }
\begin{figure}[!htbp]
    \centering
    \includegraphics[width=0.45\textwidth]{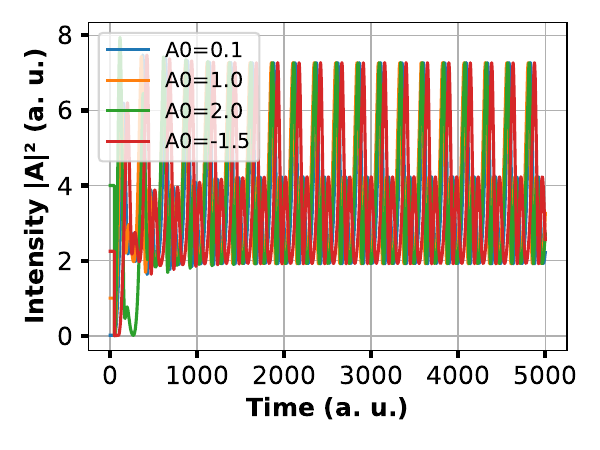}
    \caption{In the situation when the system incorporates only delay without dissipation, there are persistent oscillations, indicating that memory is retained but not stabilized.}
    \label{fig:delay_without_dissipation}
\end{figure}

In this section, the attractor convergence in the system is demonstrated, where different initial conditions evolve toward stable steady-states or stable attractors, and memory stabilization in the system, where there is retention of stable states despite perturbations, due to balance between driving, dissipation, and feedback. This is achieved through the introduction of dissipation ($\alpha$) and delayed feedback ($\kappa A(t-\Gamma)$) into equation \ref{eq:reduced_LLE} to obtain:
\begin{equation}
    \frac{dA}{dt}=-(\alpha+i\delta)A+i\gamma|A|^2 A+\kappa A(t-\Gamma)+S
\label{eq:reduced_LLE_DD}
\end{equation}

where the parameters are defined in equation \ref{eq:reduced_LLE} and $\alpha=$ dissipation, $\kappa=$ feedback strength, and $\Gamma=$ delay time steps.
The introduction of $\alpha$ and $\kappa A(t-\Gamma)$ ensures that the system exhibits stable memory with attractor formation. In the simulation, the same values for $\delta$, $\gamma$ and  $dt$ for the demonstration of nonlinear photonic memory were used. The values of $\alpha$, $\kappa$, $\Gamma$ and $S$ were set as: 1.1, 1.5, 50, and 2.5 respectively. To demonstrate the relevance of delay and dissipation within the system, the simulation is performed for 3 different scenarios: the system with delay without dissipation, the system with dissipation without delay, and the system with both delay and dissipation. For 4 different optical fields with different initial conditions, Figure ~\ref{fig:delay_without_dissipation} shows the plot of the intensity as a function of time for system case: delay without dissipation. It can be observed from the plot that, the intensity never settles to a constant value, oscillations persist throughout the entire simulation, different initial conditions continue to generate periodic dynamics, and the system does not converge to a unique steady state. So, in the absence of dissipation, delayed feedback sustains persistent oscillations, indicating that memory is retained but not stabilized.

\begin{figure}[!htbp]
    \centering
    \includegraphics[width=0.45\textwidth]{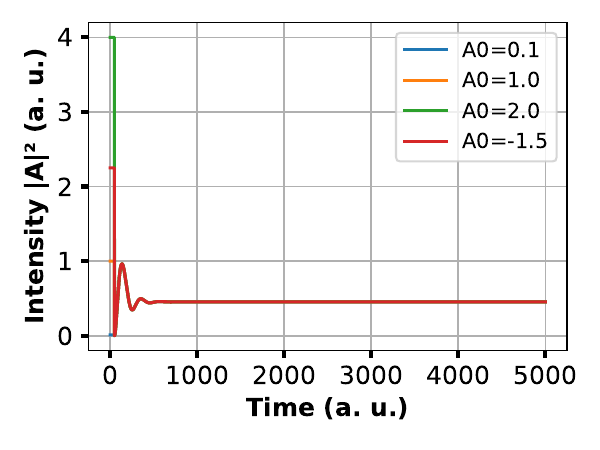}
    \caption{In the absence of delay, dissipation suppresses transient dynamics and drives the system toward a stable attractor and also, trajectories from very different initial conditions become indistinguishable after a short time}
    \label{fig:dissipation_without_delay}
\end{figure}

Figure ~\ref{fig:dissipation_without_delay} also shows the plot of the intensity as a function of time for the case of the system: dissipation without delay. It is seen from the plot that all initial conditions rapidly converge, oscillations are heavily damped, and the system reaches the same steady state. This reveals that, dissipation suppresses transient dynamics and drives the system toward a stable attractor and also, trajectories from very different initial conditions become indistinguishable after a short time.

\begin{figure}[!htbp]
    \centering
    \includegraphics[width=0.45\textwidth]{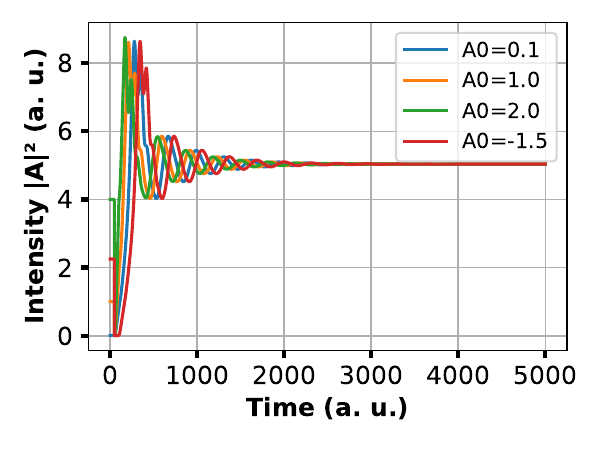}
    \caption{When both delay and dissipation are present in the system, the influence of past states (transient memory) of the system is observed, while dissipation ensures eventual convergence and stabilization}
    \label{fig:delay_plus_dissipation}
\end{figure}

Lastly, Figure ~\ref{fig:delay_plus_dissipation} shows the plot of intensity as a function of time for the system when both delay and dissipation are present. It can be seen that, initial oscillatory behavior exists, the oscillations decay gradually, all trajectories converge to the same final state, and convergence is slower than in the dissipation-only case. This demonstrates that delay introduces a memory timescale, dissipation prevents indefinite oscillation, and the system ultimately settles onto a common attractor. This clearly shows memory stabilization through attractor convergence. The convergence of trajectories originating from distinct initial conditions toward a common steady state demonstrates the existence of a stable attractor. The addition of delayed feedback extends the influence of past states (transient memory) of the system, while dissipation ensures eventual convergence and stabilization. Table ~\ref{tab:summary_system_behaviour} summarizes the behavior of the system for the 3 scenarios.

\begin{table}[h]
\centering
\caption{Summary of system behavior for 3 scenarios }
\label{tab:summary_system_behaviour}
\begin{tabular}{p{2.0cm} p{2.2cm} p{3.2cm}}
\toprule
\textbf{System with} & \textbf{Observation} & \textbf{Memory Interpretation} \\
\midrule
Delay only & Persistent oscillations & Memory retained but not stabilized \\
Dissipation only & Rapid convergence & Stable but short-lived memory \\
Delay + Dissipation & Damped oscillations followed by convergence & Stable memory with attractor formation \\
\bottomrule
\end{tabular}
\end{table}

Based on the demonstration through simulations, and from the perspective of phase-unification framework, it can be claimed that: In the dispersive regime, memory arises from phase accumulation across frequency components. In the nonlinear regime, memory emerges from intensity-dependent phase evolution and bistability. In the driven–dissipative regime, memory is stabilized through the interaction of delayed phase feedback and dissipative attractor formation.

\section{Experimental Realization of Photonic Memory in a Chirped Bragg Grating Waveguide}
\begin{figure}[!htbp]
    \centering
    \includegraphics[width=0.45\textwidth]{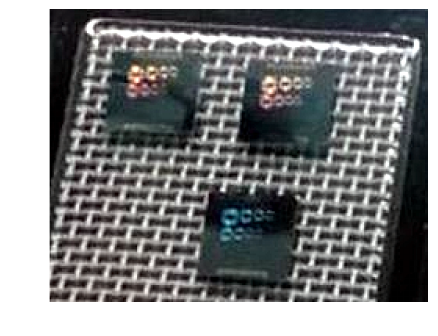}
    \caption{Photograph of three fabricated silicon photonic chips mounted on a sample holder. Each chip contains multiple spiral-shaped linearly chirped Bragg grating waveguides fabricated on a silicon-on-insulator (SOI) platform. The waveguide characterized in this work is located on one of these chips.}
    \label{fig:DUT}
\end{figure}

In this section, experimental results supporting the coexistence of dispersive, nonlinear, and driven–dissipative memory regimes within a single photonic device are presented. To the best of the author's knowledge, this phenomenon has not previously been demonstrated experimentally. Figure ~\ref{fig:DUT} shows three fabricated silicon photonic chips mounted on a sample holder. Each chip contains several spiral-shaped linearly chirped Bragg grating waveguide devices fabricated on silicon-on-insulator platform. The devices were fabricated at the Technical University of Denmark (DTU) to generate desired group delays ($\tau_{g}$), and the group delay of one of the waveguides on the chips was characterized in this work. Detailed description of how the devices were fabricated can be found in \cite{gutt2023integrated}.

The group delay experienced by an optical signal is defined as: 
\begin{equation}
    \tau_{g}=\frac{\partial \Phi}{\partial \omega}
    \label{eq:group_delay_general}
\end{equation}
where $\Phi(\omega)$ is the phase and $\omega$ is the angular frequency \cite{yorke2024fast}.

Equation ~\ref{eq:group_delay_general} follows directly from wave propagation and applies to waveguides, fibers, gratings, resonators, and cavities \cite{saleh2019fundamentals, agrawal2012nonlinear, costa1982phase}. In an optical resonator, constructive interference occurs whenever the round trip phase satisfies: $\Phi(\omega)=2 \pi m$, $m$ is an integer \cite{saleh1992fundamentals, koks2021microcavity}. For two adjacent resonances, $\Phi(\omega_{m+1})-\Phi(\omega_{m})=2 \pi$. Using equation ~\ref{eq:group_delay_general}, $\tau_{g}$ can be expressed as $\tau_{g} \approx \frac{\Delta \Phi}{\Delta \omega}=\frac{2 \pi}{2 \pi \Delta f}=\frac{1}{\Delta f}$ 
Therefore, for the special case of an optical resonator such as Michelson interferometer, the group delay for one round trip experienced by an optical pulse inside the waveguide is expressed as 
\begin{equation}
    \tau_{g}=\frac{1}{\Delta f}
\label{eq:group_delay_freq_domain}
\end{equation}
where $\Delta f$ is the frequency spacing between the peaks of the spectrum \cite{yorke2024fast, yorke2024analytical}. This means that if the reflected spectrum is measured and analyzed to obtain the peak spacing, then $\tau_{g}$ can be determined using equation ~\ref{eq:group_delay_freq_domain}.  

\begin{figure}[!htbp]
    \centering
    \includegraphics[width=0.45\textwidth]{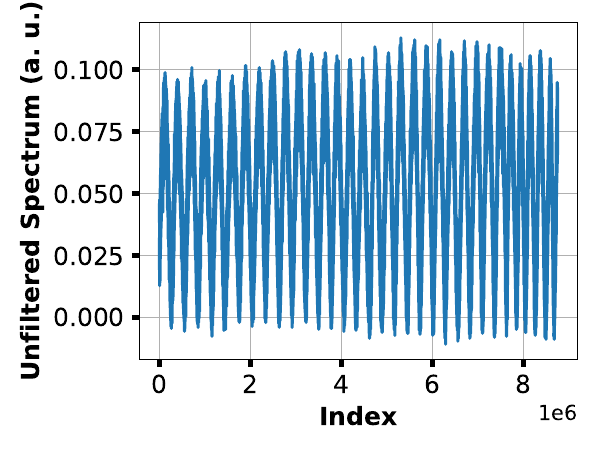}
    \caption{Zoomed-in portion of the unfiltered experimentally measured spectrum}
    \label{fig:unfiltered_spectrum}
\end{figure}

\begin{figure}[!htbp]
    \centering
    \includegraphics[width=0.45\textwidth]{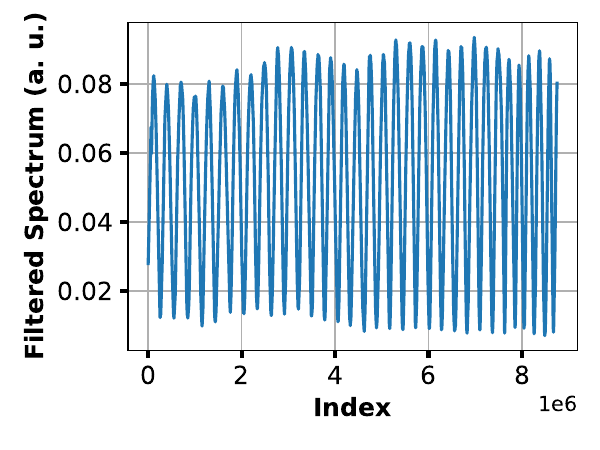}
    \caption{Zoomed-in portion of the filtered experimentally measured spectrum, which reveals decreasing peak spacing as the signal propagates due to different wavelengths  reflected at different positions within the waveguide.}
    \label{fig:filtered_spectrum}
\end{figure}

To experimentally determine the group delay of the waveguide, a measurement setup was developed at DTU to characterize the reflected spectrum of the device. The measurement principle is based on Michelson interferometry. Interferometric configurations, such as Michelson and Fabry–Pérot structures, can be realized using either uniform or chirped Bragg gratings. In these devices, the grating reflectors act as partially reflecting mirrors, generating the interference effects that form the basis of the measurement technique \cite{isaac2025innovative}. This approach exploits the interference between light reflected from the end facet of an integrated waveguide and light reflected from locations within the device during edge coupling, thereby forming an effective Michelson interferometer. The spacing of the resulting interference fringes is directly related to the group delay of the device, enabling rapid and straightforward measurement of the group delay experienced inside the waveguide \cite{yorke2024fast}. 
Detailed description of how the experiment was carried out, how the frequency axis was calibrated, how the peak spacings were determined, and the filtering strategy adopted can all be found in \cite{isaac2025innovative, yorke2024fast}. In the wavelength domain, equation ~\ref{eq:group_delay_freq_domain} can be expressed as:
\begin{equation}
    \tau_{g}=\frac{\lambda ^2}{\Delta \lambda C}
\label{eq:group_delay_lambda_domain}
\end{equation}
where $\lambda$ is the wavelength, $\Delta \lambda$ is the peak spacing in wavelength domain and $C$ is the speed of light. Figure ~\ref{fig:unfiltered_spectrum} shows the unfiltered experimentally measured reflected spectrum from the device. Figure ~\ref{fig:filtered_spectrum} also shows the filtered experimentally measured reflected spectrum, which reveals a decreasing fringe spacing as the wavelength varies. This indicates a wavelength-dependent group delay arising from the chirped Bragg grating, where different wavelengths are reflected at different positions within the waveguide.  

From Figure ~\ref{fig:filtered_spectrum}, the wavelength spacing between the peaks can be extracted and the experimentally measured wavelength-dependent group delay can be obtained using equation ~\ref{eq:group_delay_lambda_domain}.

\subsection{Interpretation of Delay as Memory}
This subsection demonstrates that the physics responsible for generating the spectrum is fundamentally a memory process. In the experiment, light enters the chirped Bragg grating waveguide, different wavelengths are reflected from different positions inside the device, the reflected signals interfere with the reflection from the waveguide facet, and this produces the oscillatory spectrum. Thus; the measured spectrum is a fingerprint of all previous reflections accumulated inside the device. In other words, the spectrum contains information about the propagation history of the optical field. In reservoir computing and neuromorphic photonics, delay is widely viewed as a memory resource. For example, delay lines provide fading memory \cite{watt2021enhancing}, delayed feedback creates temporal memory \cite{goldmann2020deep}, and reservoirs exploit delay for information retention \cite{romeira2015regenerative}. 
Also in \cite{yorke2026reconfigurable}, memory is defined as the persistence of information from previous states. In the chirped Bragg grating waveguide, a wavelength does not emerge immediately. Instead it accumulates delay: $\lambda \rightarrow \tau_{g}(\lambda)$, Thus the device retains information about the optical input over a finite time interval. This is essentially photonic fading memory. Again, in the device, instead of one delay $\tau_{g}$, there exist $\tau_{g}(\lambda)$. Thus; different wavelengths possess different memory depths. Memory ($M$) therefore is defined as
\begin{equation}
    M(\lambda) \propto \tau_{g}(\lambda)
\label{eq:photonic_memory}
\end{equation}

\begin{figure}[!htbp]
    \centering
    \includegraphics[width=0.45\textwidth]{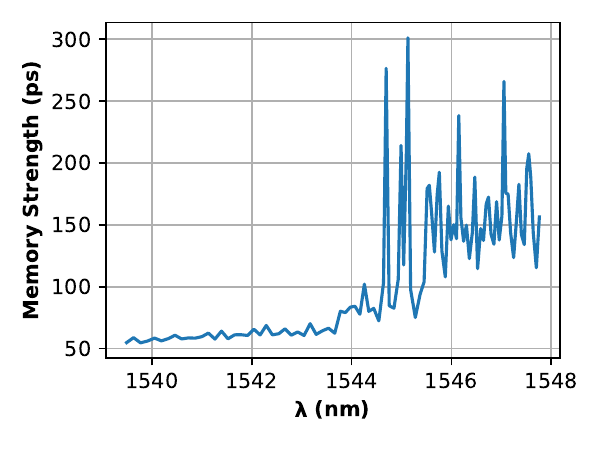}
    \caption{Experimental photonic memory as a function of wavelength obtained from the Chirped Bragg Grating Waveguide. Memory strength is defined as a quantity proportional to the experimentally measured group delay, $M(\lambda) \propto \tau_{g}(\lambda)$. Consequently, the reported memory strength is expressed in units of picoseconds.}
    \label{fig:memory_strength}
\end{figure}

Group delay is a measurable physical quantity, whilst memory is the information-retention capability, and from equation ~\ref{eq:photonic_memory}, it can be argued that $\tau_{g} \uparrow$ means the optical field spends more time inside the structure, which means $\textbf{information-retention} \uparrow$ which means $\textbf{memory-strength} \uparrow$. Figure ~\ref{fig:memory_strength} shows the experimental photonic memory distribution obtained from the chirped Bragg grating waveguide. From Figure ~\ref{fig:memory_strength}, it can be said that, the chirped Bragg grating exhibits two memory regimes. One regime exhibits relatively smooth group delay evolution, corresponding to a more stable memory state. The second regime is characterized by large group delay fluctuations, corresponding to a highly dynamic memory state. It can again be said that, the ripples correspond to local resonances, interferometric effects, or reflection interactions inside the grating. In the memory interpretation, they are memory fluctuations or local memory enhancement/suppression regions. In other words, the memory is not smooth. Certain wavelengths retain information more effectively than neighboring wavelengths.

The fast time coordinate, $\tau$ in equation ~\ref{eq:LLE} should not be confused with the experimentally measured group delay $\tau_{g}(\lambda)$. Rather, $\tau$ denotes the retarded temporal coordinate of the optical field within a moving reference frame \cite{agrawal2012nonlinear, coen2012modeling, herr2016dissipative, lugiato1987spatial}, while the wavelength-dependent group delay $\tau_{g}(\lambda)$  quantifies the wavelength-dependent propagation delay arising from the spectral phase. Through chromatic dispersion, the measured group delay distribution governs how spectral components are mapped onto the fast time coordinate \cite{agrawal2012fiber}. Consequently, within the proposed framework, the experimentally measured group delay distribution provides the physical basis for interpreting temporal photonic memory.

\section{Unified memory interpretation of the experimentally realized photonic memory}
From equation ~\ref{eq:photonic_memory}, this work proposes that $M(\lambda) \propto \tau_{g}(\lambda)$, this means fluctuations in group delay $Var(\tau)$ imply fluctuations in memory $Var(M)$. Thus; large delay fluctuations $\rightarrow$ highly variable memory and small delay fluctuations $\rightarrow$ stable memory. The linear region in Figure ~\ref{fig:memory_strength} corresponds to the linear memory regime, characterized by $\frac{d \tau}{d \lambda} \approx small$ and $Var(\tau) \approx smaill$. In this regime, there is	predictable memory, stable storage, and near-linear response. The nonlinear region in Figure ~\ref{fig:memory_strength} corresponds to the nonlinear memory regime, characterized by $\frac{d \tau}{d \lambda}$ changing rapidly and $Var(\tau)$ is large. In this regime, there is memory fluctuations, enhanced sensitivity and nonlinear processing behavior. The repeated reflections inside the chirped Bragg grating naturally create
energy injection from the input field, redistribution through interference, and gradual loss through propagation and scattering. So the system is not merely storing information. It is continuously doing: $Drive \rightarrow Store \rightarrow Release\rightarrow Dissipate$. This is exactly the intuition behind a driven-dissipative memory system.


\section{Discussion}
This work has established a unified framework in which photonic memory emerges through three distinct yet interconnected dynamical phases: the dispersive, nonlinear, and driven–dissipative regimes. Rather than representing independent physical mechanisms, these regimes are shown to arise from the evolution of the optical phase, with transitions governed by the interplay between dispersion, Kerr nonlinearity, and dissipative feedback. This perspective provides a common physical description for memory formation across a broad class of photonic systems and suggests that photonic memory should be viewed as a dynamical phase of the optical field rather than as a property unique to a specific device architecture.
The theoretical analysis demonstrates that each regime possesses a characteristic memory mechanism. In the dispersive regime, memory originates from frequency-dependent phase accumulation, which generates temporal correlations between successive states of the optical field. The nonlinear regime extends this behavior through Kerr-induced intensity-dependent phase evolution, enabling nonlinear feedback, bistability, and hysteresis, thereby introducing history-dependent memory. Finally, the driven–dissipative regime incorporates dissipation and delayed feedback, producing attractor convergence and memory stabilization through the balance between information retention and energy loss. Together, these results establish a continuous progression from transient temporal memory to nonlinear state-dependent memory and ultimately to stable attractor-based memory.
The experimental results provide strong support for the proposed framework. Measurements of the group delay distribution reveal two distinct operational regimes within the same chirped Bragg grating device. A smooth group delay response exhibits relatively small delay variations and is consistent with a stable linear memory regime dominated by dispersive phase evolution. In contrast, a strongly fluctuating group delay response exhibits significantly larger delay variations, indicating the onset of nonlinear phase dynamics and a corresponding nonlinear memory response. The observation of these distinct operating regimes within a single device demonstrates that the transition from linear to nonlinear photonic memory can be realized experimentally without modifying the underlying device architecture.
Most importantly, the experimentally reconstructed memory distribution shown in Figure ~\ref{fig:memory_strength} provides evidence that dispersive, nonlinear, and driven–dissipative memory characteristics can coexist within the same photonic platform. The measured group delay distributions capture the dispersive and nonlinear operating regimes, while the resonant cavity formed by the chirped Bragg grating inherently provides the driven–dissipative environment through repeated optical feedback and intrinsic loss. Collectively, these observations support the proposed unified description of photonic memory as a phase-driven dynamical phenomenon. Within the proposed unified memory framework, the measured group delay is interpreted as the photonic memory strength according to $M(\lambda) \propto \tau_{g}(\lambda)$. The measured group delay (memory) window extends from approximately 50 ps to 300 ps, corresponding to a memory window of approximately 250 ps.
The proposed framework has broader implications for neuromorphic photonics and photonic information processing. By identifying optical phase evolution as the common physical quantity underlying memory formation, the framework establishes a unified foundation for designing photonic systems capable of simultaneously exploiting transient memory, nonlinear computation, and attractor stabilization. Such systems may enable compact and energy-efficient photonic hardware that integrates memory and computation within a single device, providing new opportunities for reservoir computing, photonic neural networks, and future integrated neuromorphic processors.

\section{Conclusion}
This work has presented a unified framework for photonic memory by demonstrating that memory can exist in three interconnected dynamical phases: dispersive, nonlinear, and driven–dissipative. Theoretical analysis showed that transitions between these regimes are governed by optical phase evolution, Kerr nonlinearity, and the balance between delayed feedback and dissipation. Experimental measurements of the memory distribution further revealed distinct linear and nonlinear memory responses within a chirped Bragg grating, while the reconstructed memory distribution provides evidence supporting the coexistence of all three memory regimes within a single integrated photonic device. These findings establish optical phase as the common physical quantity underlying photonic memory formation and provide a unified perspective that bridges wave propagation, nonlinear dynamics, and driven–dissipative systems.
Extending the proposed framework to programmable integrated photonic platforms may enable compact architectures capable of simultaneously performing memory, nonlinear computation, and attractor-based information processing, thereby providing a foundation for next generation photonic neural networks, reservoir computing systems, and energy-efficient optical artificial intelligence hardware.

\section{Data Availability Statement}
The data supporting the findings of this study are available from the author upon reasonable request, subject to reasonable research and intellectual property considerations.

\section*{Acknowledgment}
I sincerely thank my parents for their unwavering prayers, encouragement, and support throughout this research journey. I also gratefully acknowledge Dr. Peter David Girouard for providing me with a strong foundation in academic research and for his guidance in shaping my research career. Finally, my heartfelt gratitude goes to my colleagues who have been encouraging me to continue my independent research.

\ifCLASSOPTIONcaptionsoff
  \newpage
\fi

\bibliographystyle{IEEEtran}
\bibliography{references}

@article{goi2020perspective,
  title={Perspective on photonic memristive neuromorphic computing},
  author={Goi, Elena and Zhang, Qiming and Chen, Xi and Luan, Haitao and Gu, Min},
  journal={PhotoniX},
  volume={1},
  number={1},
  pages={3},
  year={2020},
  publisher={Springer}
}

@article{farmakidis2024integrated,
  title={Integrated photonic neuromorphic computing: opportunities and challenges},
  author={Farmakidis, Nikolaos and Dong, Bowei and Bhaskaran, Harish},
  journal={Nature Reviews Electrical Engineering},
  volume={1},
  number={6},
  pages={358--373},
  year={2024},
  publisher={Nature Publishing Group UK London}
}

@article{van2017advances,
  title={Advances in photonic reservoir computing},
  author={Van der Sande, Guy and Brunner, Daniel and Soriano, Miguel C},
  journal={Nanophotonics},
  volume={6},
  number={3},
  pages={561--576},
  year={2017},
  publisher={De Gruyter}
}

@article{yorke2026reconfigurable,
  title={Reconfigurable Nonlinear Photonic Networks for In-Situ Learning and Memory Formation via Driven-Dissipative Dynamics},
  author={Yorke, Isaac},
  journal={arXiv preprint arXiv:2605.19911},
  year={2026}
}

@article{patterson2022computer,
  title={Computer organization and Design},
  author={Patterson, David A and Hennessy, John L},
  journal={The Hardware/Soft},
  year={2022}
}

@book{prucnal2017neuromorphic,
  title={Neuromorphic photonics},
  author={Prucnal, Paul R and Shastri, Bhavin J},
  year={2017},
  publisher={CRC press}
}

@book{gibbs2012optical,
  title={Optical bistability: controlling light with light},
  author={Gibbs, Hyatt},
  year={2012},
  publisher={Elsevier}
}

@article{lukovsevivcius2009reservoir,
  title={Reservoir computing approaches to recurrent neural network training},
  author={Luko{\v{s}}evi{\v{c}}ius, Mantas and Jaeger, Herbert},
  journal={Computer science review},
  volume={3},
  number={3},
  pages={127--149},
  year={2009},
  publisher={Elsevier}
}

@article{paparelle2026experimental,
  title={Experimental memory control in continuous-variable optical quantum reservoir computing},
  author={Paparelle, Iris and Henaff, Johan and Garcia-Beni, Jorge and Gillet, Emilie and Montesinos, Daniel and Giorgi, Gian Luca and Soriano, Miguel C and Zambrini, Roberta and Parigi, Valentina},
  journal={Nature Photonics},
  pages={1--8},
  year={2026},
  publisher={Nature Publishing Group UK London}
}

@inproceedings{castro2024memory,
  title={Memory capacity analysis of time-delay reservoir computing based on silicon microring resonator nonlinearities},
  author={Castro, Bernard J Giron and Peucheret, Christophe and Da Ros, Francesco},
  booktitle={Machine Learning in Photonics},
  volume={13017},
  pages={115--126},
  year={2024},
  organization={SPIE}
}

@article{labay2023quantum,
  title={Quantum associative memory with a single driven-dissipative nonlinear oscillator},
  author={Labay-Mora, Adri{\`a} and Zambrini, Roberta and Giorgi, Gian Luca},
  journal={Physical Review Letters},
  volume={130},
  number={19},
  pages={190602},
  year={2023},
  publisher={APS}
}

@article{bea1991fundamentals,
  title={Fundamentals of photonics},
  author={BEA, SALEH and Teich, MC},
  journal={Wiley},
  pages={313},
  year={1991}
}

@book{agrawal2012nonlinear,
  title={Nonlinear fiber optics},
  author={Agrawal, Govind},
  year={2012},
  publisher={Elsevier}
}

@book{ablowitz1981solitons,
  title={Solitons and the inverse scattering transform},
  author={Ablowitz, Mark J and Segur, Harvey},
  year={1981},
  publisher={SIAM}
}

@book{saleh2019fundamentals,
  title={Fundamentals of photonics, 2 volume set},
  author={Saleh, Bahaa EA and Teich, Malvin Carl},
  year={2019},
  publisher={john Wiley \& sons}
}

@article{lugiato1987spatial,
  title={Spatial dissipative structures in passive optical systems},
  author={Lugiato, Luigi A and Lefever, Ren{\'e}},
  journal={Physical review letters},
  volume={58},
  number={21},
  pages={2209},
  year={1987},
  publisher={APS}
}

@article{asuero2006correlation,
  title={The correlation coefficient: An overview},
  author={Asuero, Agustin Garcia and Sayago, Ana and Gonz{\'a}lez, AG},
  journal={Critical reviews in analytical chemistry},
  volume={36},
  number={1},
  pages={41--59},
  year={2006},
  publisher={Taylor \& Francis}
}

@article{loures2015contribution,
  title={Contribution of third-harmonic and negative-frequency polarization fields to self-phase modulation in nonlinear media},
  author={Loures, Cristian Redondo and Armaroli, Andrea and Biancalana, Fabio},
  journal={Optics Letters},
  volume={40},
  number={4},
  pages={613--616},
  year={2015},
  publisher={Optical Society of America}
}

@article{mezache2026nonlinear,
  title={Nonlinear Schr{\"o}dinger equation solutions for bessel pulse propagation in chiral kerr media with multiphoton absorption},
  author={Mezache, Zinelabiddine and Berka, Mohammed},
  journal={The European Physical Journal Plus},
  volume={141},
  number={3},
  pages={225},
  year={2026},
  publisher={Springer}
}

@article{maurya2012pattern,
  title={Pattern formation in a passive incoherent ring resonator system based on nonlinear material with the self-focusing non-instantaneous Kerr response},
  author={Maurya, MK and Yadav, RA},
  journal={Optics \& Laser Technology},
  volume={44},
  number={3},
  pages={505--513},
  year={2012},
  publisher={Elsevier}
}

@book{lidorikis1999wave,
  title={Wave propagation in ordered, disordered, and nonlinear photonic band gap materials},
  author={Lidorikis, Elefterios Efstathiou},
  year={1999},
  publisher={Iowa State University}
}

@inproceedings{priem2003optical,
  title={Optical phase shifting with 1D Kerr-nonlinear resonators},
  author={Priem, Gino and Morthier, Geert and Baets, Roel},
  booktitle={Proceedings of IEEE/LEOS 2003, Proceedings of the Eight Annual Symposium of the IEEE/LEOS Benelux Chapter},
  pages={241--244},
  year={2003}
}

@article{petracek2014simulation,
  title={Simulation of self-pulsing in Kerr-nonlinear coupled ring resonators},
  author={Petr{\'a}cˇek, Jirˇ{\'\i} and Ek{\c{s}}io{\u{g}}lu, Yasa and Sterkhova, Anna},
  journal={Optics Communications},
  volume={318},
  pages={147--151},
  year={2014},
  publisher={Elsevier}
}

@book{isaac2025innovative,
  title={Innovative photonics-based devices and measurement techniques for modern optical communications},
  author={Isaac, Yorke},
  year={2025},
  publisher={Universit{\`a} degli Studi di Parma. Dipartimento di Ingegneria e architettura}
}

@book{gutt2023integrated,
  title={Integrated Optical Time Lenses for Passive Optical Networks: For Generation Of On-Chip Optical Fourier Transforms},
  author={Gutt, Lars Emil},
  year={2023},
  publisher={Technical University of Denmark}
}

@inproceedings{yorke2024fast,
  title={Fast Method for the Measurement of Dispersion of Integrated Waveguides by Utilizing Michelson Interferometry Effects},
  author={Yorke, Isaac and Gutt, Lars Emil and Girouard, Peter David and Galili, Michael},
  booktitle={Physical Sciences Forum},
  volume={10},
  number={1},
  pages={4},
  year={2024},
  organization={MDPI}
}

@inproceedings{yorke2024analytical,
  title={Analytical model for dispersion measurement in integrated waveguides using michelson interferometry effects},
  author={Yorke, Isaac and Girouard, Peter David and Galili, Michael},
  booktitle={EPJ Web of Conferences},
  volume={309},
  pages={03006},
  year={2024},
  organization={EDP Sciences}
}

@article{watt2021enhancing,
  title={Enhancing computational performance of a spin-wave reservoir computer with input synchronization},
  author={Watt, Stuart and Kostylev, Mikhail and Ustinov, Alexey B},
  journal={Journal of Applied Physics},
  volume={129},
  number={4},
  year={2021},
  publisher={AIP Publishing}
}

@article{goldmann2020deep,
  title={Deep time-delay reservoir computing: Dynamics and memory capacity},
  author={Goldmann, Mirko and K{\"o}ster, Felix and L{\"u}dge, Kathy and Yanchuk, Serhiy},
  journal={Chaos: An Interdisciplinary Journal of Nonlinear Science},
  volume={30},
  number={9},
  year={2020},
  publisher={AIP Publishing}
}

@article{romeira2015regenerative,
  title={Regenerative memory in time-delayed neuromorphic photonic systems},
  author={Romeira, B and Av{\'o}, R and Figueiredo, Jos{\'e} ML and Barland, S and Javaloyes, J},
  journal={arXiv preprint arXiv:1503.07781},
  year={2015}
}

@article{costa1982phase,
  title={Phase shift technique for the measurement of chromatic dispersion in optical fibers using LED's},
  author={Costa, Bruno and Mazzoni, Daniele and Puleo, Mario and Vezzoni, Emilio},
  journal={IEEE Transactions on Microwave Theory and Techniques},
  volume={30},
  number={10},
  pages={1497--1503},
  year={1982},
  publisher={IEEE}
}

@misc{saleh1992fundamentals,
  title={Fundamentals of photonics},
  author={Saleh, Bahaa and Teich, Malvin and Slusher, Richard E},
  year={1992},
  publisher={American Institute of Physics}
}

@article{koks2021microcavity,
  title={Microcavity resonance condition, quality factor, and mode volume are determined by different penetration depths},
  author={Koks, Corn{\'e} and Van Exter, MP},
  journal={Optics Express},
  volume={29},
  number={5},
  pages={6879--6889},
  year={2021},
  publisher={Optical Society of America}
}

@article{chembo2013spatiotemporal,
  title={Spatiotemporal Lugiato-Lefever formalism for Kerr-comb generation in whispering-gallery-mode resonators},
  author={Chembo, Yanne K and Menyuk, Curtis R},
  journal={Physical Review A—Atomic, Molecular, and Optical Physics},
  volume={87},
  number={5},
  pages={053852},
  year={2013},
  publisher={APS}
}

@article{coen2012modeling,
  title={Modeling of octave-spanning Kerr frequency combs using a generalized mean-field Lugiato--Lefever model},
  author={Coen, St{\'e}phane and Randle, Hamish G and Sylvestre, Thibaut and Erkintalo, Miro},
  journal={Optics letters},
  volume={38},
  number={1},
  pages={37--39},
  year={2012},
  publisher={Optical Society of America}
}

@article{herr2016dissipative,
  title={Dissipative Kerr solitons in optical microresonators},
  author={Herr, Tobias and Gorodetsky, Michael L and Kippenberg, Tobias J},
  journal={Nonlinear optical cavity dynamics: from microresonators to fiber lasers},
  pages={129--162},
  year={2016},
  publisher={Wiley Online Library}
}

@book{agrawal2012fiber,
  title={Fiber-optic communication systems},
  author={Agrawal, Govind P},
  year={2012},
  publisher={John Wiley \& Sons}
}
\end{document}